\documentclass[12pt]{article}
\usepackage{float}
\usepackage{amssymb}
\usepackage{amsthm}
\usepackage{amsmath}
\usepackage{latexsym}
\usepackage{epsfig}
\usepackage{graphicx}
\usepackage{wasysym}
\usepackage{threeparttable}
\usepackage{natbib}
\usepackage{color}
\usepackage{epstopdf}
\usepackage{caption}
\usepackage{subcaption}

\newcommand{\mbf}[1]{\mathbf{#1}}

\usepackage[breaklinks]{hyperref}

\usepackage{enumitem}

\def\bb{\mathbf{b}}

\def\bh{\mathbf{h}}

\def\bs{\mathbf{s}}
\def\bu{\mathbf{u}}

\def\bB{\mathbf{B}}

\def\bH{\mathbf{H}}
\def\bI{\mathbf{I}}
\def\bK{\mathbf{K}}
\def\bM{\mathbf{M}}

\def\bY{\mathbf{Y}}

\newcommand{\bfeta}{\mbox{\boldmath $\eta$}}

\newcommand{\bfzeta}{\mbox{\boldmath $\zeta$}}

\newcommand{\bfPhi}{\mbox{\boldmath $\Phi$}}
\newcommand{\bfPsi}{\mbox{\boldmath $\Psi$}}

\newcommand{\bfpsi}{\mbox{\boldmath $\psi$}}
\newcommand{\bfmu}{\mbox{\boldmath $\mu$}}

\newcommand{\bfnu}{\mbox{\boldmath $\nu$}}

\newcommand{\bfLambda}{\mbox{\boldmath $\Lambda$}}
\newcommand{\bfSigma}{\mbox{\boldmath $\Sigma$}}
\newcommand{\var}{\textrm{var}}
\newcommand{\cov}{\textrm{cov}}

\usepackage{color}

\def\boxit#1{\vbox{\hrule\hbox{\vrule\kern6pt
          \vbox{\kern6pt#1\kern6pt}\kern6pt\vrule}\hrule}}

\def\var{\hbox{var}}
\def\cov{\hbox{cov}}

\def\bse{\begin{eqnarray*}}
\def\ese{\end{eqnarray*}}
\def\be{\begin{eqnarray}}
\def\ee{\end{eqnarray}}
\def\bq{\begin{equation}}
\def\eq{\end{equation}}
\def\bse{\begin{eqnarray*}}
\def\ese{\end{eqnarray*}}

\usepackage[ruled]{algorithm2e}
\usepackage{mathptmx}      
\usepackage{bm}

 {\begin{list}{}%
         {\setlength{\leftmargin}{#1}}%
         \item[]%
 }
 {\end{list}}
\usepackage{setspace}

\begin{document}
\thispagestyle{empty} \baselineskip=28pt

\begin{center}
{\LARGE{\bf Spatio-Temporal Change of Support with Application to American Community Survey Multi-Year Period Estimates}}
\end{center}

\baselineskip=12pt

\vskip 2mm
\begin{center}
Jonathan R. Bradley\footnote{(\baselineskip=10pt to whom correspondence should be addressed) Department of Statistics, University of Missouri, 146 Middlebush Hall, Columbia, MO 65211, bradleyjr@missouri.edu},
Christopher K. Wikle\footnote{\baselineskip=10pt  Department of Statistics, University of Missouri, 146 Middlebush Hall, Columbia, MO 65211-6100},
Scott H. Holan$^2$
\end{center}
%
%
%
%
\vskip 4mm

\begin{center}
\large{{\bf Abstract}}
\end{center}
We present hierarchical Bayesian methodology to perform spatio-temporal change of support (COS) for survey data with Gaussian sampling errors.  This methodology is motivated by the American Community Survey (ACS), which is an ongoing survey administered by the U.S. Census Bureau that provides timely information on several key demographic variables. The ACS has published 1-year, 3-year, and 5-year period-estimates, and margins of errors, for demographic and socio-economic variables recorded over predefined geographies.  The spatio-temporal COS methodology considered here provides data users with a way to estimate ACS variables on customized geographies and time periods, while accounting for sampling errors. Additionally, 3-year ACS period estimates are to be discontinued, and this methodology can provide predictions of ACS variables for 3-year periods given the available period estimates.  The methodology is based on a spatio-temporal mixed effects model with a low-dimensional spatio-temporal basis function representation, which provides multi-resolution estimates through basis function aggregation in space and time.  This methodology includes a novel parameterization that uses a target dynamical process and recently proposed parsimonious Moran's I propagator structures.  Our approach is demonstrated through two applications using public-use ACS estimates, and is shown to produce good predictions on a holdout set of 3-year period estimates.

\baselineskip=12pt

%
%
%

\baselineskip=12pt
\par\vfill\noindent
\par\medskip\noindent
\clearpage\pagebreak\newpage \pagenumbering{arabic}
\baselineskip=24pt
\section{Introduction}

The American Community Survey (ACS) is an ongoing survey that 
releases data annually, providing communities with the current 
information needed to plan investments and services. The ACS was designed 
to produce reliable annually updated estimates for topics 
only previously available once every decade from the decennial census 
long-form, such as detailed demographic housing and 
socioeconomic topics. The structure of the publicly released ACS data is unique, consisting of rolling 
multi-year estimates (MYEs). 
In August 2006, the U.S. Census Bureau released its first set of one-year estimates 
for areas with populations greater than 65,000.  Subsequently, 
in December 2008, the first set of three-year 
estimates were released for areas with populations greater than 20,000.  
Completion of three types of releases culminated in 2010 
with the release of the five-year estimates for all standard 
tabulation areas including census tracts and block groups.  
It was recently announced, however, that the 3-year MYEs are to be discontinued {in the beginning of the 2016 fiscal year \citep{3yearperiod}. For additional details on the structure of ACS, see \citet{acsSummary} and \citet{torrieri}.}

The shift from the decennial census long-form data to using 
MYEs from the ACS offers unique challenges and opportunities for data practitioners. One of these is to transform the survey estimates from one geography and/or time period to another in a manner that can account for sampling uncertainty (i.e., allowing for custom geographies and/or time periods). In general, procedures that allow one to perform statistical inference on a spatio-temporal support that differs from the support of the data (either in space, time, or space and time) is referred to as change of support (COS). In this setting, we let the survey data's spatio-temporal support (e.g., census tracts and a 5-year time period) be denoted as the source support, and let the support of practical interest be called the target support (which might also be a standard census geography). In particular, we are interested in the problem where either the spatial and/or temporal support differs between the source and target. Allowing ACS data-users to obtain estimates and conduct inference on user-defined spatio-temporal supports has recently been identified as an important problem by a National Academy of Sciences (NAS) panel \citep{nationacad}.  

{Clearly, the introduction and removal of 3-year period estimates has produced a need for spatio-temporal COS among the ACS data-user community. {Although there is interest in this particular COS problem}, there are broader implications of spatio-temporal COS that are of interest to federal statisticians and ACS data-users. For example, \citet{tucker} proposed methodology for comparing trends between counties that do not have compatible MYEs. A method of spatio-temporal COS would allow one to readily compare (time) trends across different geographies, with potentially increased precision. Another example can be found in \citet{bradleyCOS}, where New York City's Department of City Planning is interested in producing estimates of socio-economic variables on community districts (a geography not made available by ACS). {Furthermore, understanding the multi-scale structure of ACS period estimates in space \citep[e.g., see][among others]{Siordia1,Bazuin,Siordia2,Siordia3,Siordia4} and/or time \citep[e.g., see][among others]{tucker,acsTuck,acsTuck2} is a consistent theme in the ACS literature.} These reoccurring problems suggest that spatio-temporal COS would greatly enhance the utility of ACS period estimates. Thus, our primary goal is to introduce a method to perform spatio-temporal COS for survey data, with sampling errors that are reasonably modeled as Gaussian random variables.}

In the geographical sciences, a standard approach to the spatial-only COS problem is known as {\it simple areal interpolation}, where values are imputed proportionally based on the areas of each of the target areal units \citep[e.g.,][]{flowerden,green3,rogerskill,green2,green}. Although such procedures are easy to implement, measures of uncertainty are not readily available, which limits their usefulness in situations with substantial measurement/sampling uncertainty.

Spatial statistics is an avenue for spatial COS that takes into account uncertainty. For reviews, see \citet{Gotway}, \citet{cressie-wikle-book}, and \citet{banerjee-etal-2004}. There are two general methodological approaches currently used for spatial COS. The first method involves defining a spatial process at the point level, and then integrating the process to the desired target support. For an example of this ``bottom-up'' strategy see \citet{WileBerliner} where they consider a hierarchical Bayesian approach for Gaussian data that includes simple areal interpolation as a special case; also see \citet{bradleyCOS} who consider a hierarchical spatial Poisson COS methodology. The second general modeling framework for spatial COS starts by defining parameters for each set in a partitioning between the source and target support. For examples of this ``top-down'' strategy see \citet{mugglin98} and \citet{mugglin99}, where they use a hierarchical Bayesian approach based on a Poisson data model. In general, both of these approaches give similar results; see \citet{gelfandCOSreview} for an explicit comparison.

We choose to specify the latent process on a point-level spatial support and integrate to any desired target support (i.e., the ``bottom-up'' method for spatial COS). This allows one to avoid computationally expensive Bayesian simulation every time a new target support of interest is introduced \citep[e.g., see the discussion in][]{bradleyWiley,bradleyCAGE}. This flexibility is especially valuable for our application, since ACS users may continually propose new target supports.

Although the literature for spatial COS is mature, very little statistical work has been done on {\it spatio-temporal} COS.  We focus on such a methodology for data that are reasonably modeled as Gaussian.  The methodology explicitly accounts for sampling uncertainty in the survey estimates, as well as differing geographies and MYEs.  We utilize a Bayesian hierarchical modeling framework with a key assumption that the important spatio-temporal variability in the latent spatio-temporal variable of interest can be efficiently modeled in terms of a relatively low-rank spatio-temporal basis expansion, with associated random expansion coefficients.  This modeling approach allows for the COS methodology to be applied to high-dimensional datasets.  Although various basis expansion approaches have become the norm in spatial statistics over the last few years, both for spatial processes and spatio-temporal processes \citep[e.g., see][for an overview]{cressie-wikle-book}, these models have not typically considered joint spatio-temporal basis function expansions.  The joint spatio-temporal basis functions are important in our methodology as they allow for simple aggregation across spatial scales and time.

  In addition, we present a novel parameterization of the random effects dependence structure that makes use of a ``target'' dynamical spatio-temporal process and what we have termed the {\it Moran's I propagator} \citep[see][]{bradleyMSTM}.  This is important as it respects the fact that there is an underlying dynamical process, but recognizes that we only need be concerned with the implied marginal dependence suggested by that process.  This parameterization is further characterized by its extreme parsimony, thus providing a fairly low-dimensional parameter space for our Bayesian hierarchical model.  We illustrate this methodology by considering the MYEs for ACS median household income and predict the 3-year MYE for 2013 (which was held out of our analysis) to provide a demonstration that our methodology provides a principled approach for predicting the discontinued 3-year MYEs for ACS data. {Furthermore, a demonstration of simultaneous spatial and temporal COS is given, where 1-, 3-, and 5-year period estimates and counties are used as the source support, and 1-, 2-, 3-, and 4- year period estimates and ``American Indian area/Alaska native area/Hawaiian home lands'' are defined as the target support.}

The remainder of this paper proceeds as follows. The spatio-temporal COS methodology is outlined in Section 2, with the ACS example presented in Section 3.  We conclude in Section 4 with a brief discussion.

\section{Methodology}
The data of interest rely crucially on the sample design as well as the direct estimates and their estimated variances.  
For the sake of generality we present our approach in terms of MYEs. Note, for ease of notation we refer to a 1-year period estimate as a MYE. Let $Z_t^{(\ell)}(A)$ represent the ACS data quantity of  interest for the $\ell$-th time period ($\ell=1,3,5$), the $t$-th year, and for spatial region $A \in D_{t,A}^{(\ell)}$, where the set $D_{t,A}^{(\ell)}$ is a collection of areal units that can be different depending on the period and time.  We then specify the following data model
\begin{align}\label{additive}
Z_{t}^{(\ell)}(A)=Y_{t}^{(\ell)}(A)+\epsilon_{t}^{(\ell)}(A);\hspace{5pt} A \in D_{t,A}^{(\ell)},\hspace{5pt}t = T_{L},\ldots,T_{U},\hspace{5pt}\ell = 1,3,5,
\end{align}
where $Y_{t}^{(\ell)}(A)$ is considered to be the ``true'' (but unknown) latent variable of interest at time period $\ell$, time $t$, and spatial region $A$,
and the ``survey error'' $\epsilon_{t}^{(\ell)}(A)$ is assumed to be an independent and identically distributed (i.i.d.) Gaussian random variable with mean zero and variance $\sigma_{t,\ell}^{2}(A)$. These variances are assumed known and computed from ACS estimates of the margin of error associated with $\{Z_{t}^{(\ell)}(A)\}$. 

\subsection{Process Model} 
One of the challenges with spatio-temporal COS is to formulate a flexible spatio-temporal model for the latent process, $Y_{t}^{(\ell)}(A)$.  It is intuitive, and customary in geostatistics, to consider COS from the perspective of aggregating a point-scale latent process \citep[e.g., see][Section 4.1.3]{cressie-wikle-book}.  That is, for any spatial region $S \subset D_S \in \mathbb{R}^d$ and period $\ell$, we can define the discrete-time, continuous-space, spatio-temporal aggregate as
\begin{equation}
Y^{(\ell)}_t(S) \equiv \sum_{k=t-\ell -1}^t \frac{\omega_t}{|S|} \int_{D_S} Y(\bs;k) d\bs, \; |S| > 0, 
\label{eq:STCOS}
\end{equation}
where $|S| = \int_{D_S} d\bs$, $\omega_t$ are temporal weights, and $Y(\bs;k)$ is a continuous-space, discrete-time spatio-temporal process defined at spatial locations $\bs \in D_S$ and times $k \in \{T_L,\ldots,T_U\}$.

The aforementioned methodology relies on the specification of the ``point-level'' spatio-temporal process.  In particular, consider the point-level spatial-temporal representation
\begin{equation}
Y(\bu;k) = \delta(\bu) + \sum_{j=1}^\infty {\psi}_j(\bu;k)\; \eta_j, \; \bu \in D_s, k \in D_t,
\label{eq:CSDTmodel}
\end{equation}
where $\delta(\bu)$ is a large-scale spatial trend term, $\{\psi_j(\bu;k)\}_{j=1}^{\infty}$ corresponds to a pre-specified set of basis functions indexed in space ($\bu$) and time ($k$), and $\{\eta_j\}$ are the random expansion coefficients associated with each basis function, assumed to be mean zero Gaussian and potentially dependent (see below). At this point, we leave the specification of the basis functions general, but present a specific example in Section 2.2.  

Upon substitution of (\ref{eq:CSDTmodel}) into (\ref{eq:STCOS}), we have the following model for $Y_t^{(\ell)}(A)$:
\begin{eqnarray}
Y_t^{(\ell)}(A) & = & \frac{1}{|A|} \int_A \delta(\bu)d\bu + \frac{1}{\ell \; |A|} \sum_{k=t-\ell+1}^t \int_A \sum_{j=1}^{\infty} {\psi}_j(\bu;k) \eta_j \; d\bu, \nonumber \\
& = & \mu(A) + \frac{1}{\ell \; }\sum_{k=t-\ell+1}^t  \sum_{j=1}^{r} \psi_j(A;k) \eta_j + \xi_t^{(\ell)}(A),
\label{eq:Yagg}
\end{eqnarray}
where we have assumed $\omega_t = 1/\ell$, defined $\mu(A) \equiv \frac{1}{|A|} \int_A \delta(\bu) d\bu$, $\; \psi_j(A;k)\equiv\frac{1}{|A|} \int_A \psi_j(\bu;k) d\bu$, and we have assumed that, given a sufficiently large $r$ and upon integration, the finite truncation of the basis expansion is a reasonable approximation. That is, we have assumed
\begin{equation*}\label{truncresid}
	\sum_{j = r+1}^{\infty}\frac{1}{\ell}\sum_{k=t-\ell + 1}^{t}\frac{1}{|A|}\int_{A}\psi_{j}(\bu;k)d\bu\hspace{5pt}\eta_{j}\approx 0; \hspace{5pt}A \subset \mathbb{R}^{d}, t = T_{L},\ldots,T_{U}, \ell = 1,3,5.
\end{equation*}
To account for this truncation and potential model misspecification, we include the ``fine scale'' error term $\{\xi_t^{(\ell)}(A)\}$, which are assumed to be i.i.d. Gaussian random variables with mean zero and variance $\sigma^2_\xi$.  Note that we are assuming this truncation error applies at the aggregate spatial scale.  

In practice, it is convenient to define a spatial support at which parameters are assumed to be constant at lower resolutions. Let $D_{B}\equiv \{B_{i}: i = 1,\ldots,n_{B}\}$ be this pre-defined fine-level spatial support consisting of disjoint areal units. For example, in our application (Section 3), we let $D_{B}$ consist of every US county as defined in 2013 (note that county definitions change from year-to-year). Now, let $\delta(\bu) = \mu_{i}$ for any $\bu \in \mathbb{R}^{d}$ such that $\bu \in B_{i}$, $B_{i}$ is the $i$-th areal unit in $D_{B}$, $\mu_{i}\in \mathbb{R}$ is unknown, and $i = 1,\ldots,n_{B}$.  This implies that the spatial trend term is constant within each of the $i = 1,\ldots,n_{B}$ areal units in $D_{B}$ (with the respective value $\mu_{i}$). If we partition $A\in D_{t,A}^{(\ell)}$ into its potential overlap with all $B \in D_{B}$ then
	\begin{align}\label{partition}
	\mu(A) &= \frac{1}{|A|}\sum_{i = 1}^{{n_{B}}}\underset{A \cap {B_{i}}}{\int}\delta(\bu)d\bu=\frac{1}{|A|}\sum_{i = 1}^{{n_{B}}}{\mu_{i}}\underset{A \cap {B_{i}}}{\int}1 \; d\bu =\sum_{i = 1}^{{n_{B}}}\mu_{i}\frac{|A\cap {B_{i}}|}{|{A}|}.
	\end{align}
	\noindent
	Letting $h(A,i) \equiv |A \cap {B_{i}}|/|{A}|$ and defining the ${n_{B}}$-dimensional vectors $\textbf{h}(A)\equiv(h(A,1),$ $\ldots,h(A,{n_{B}}))^{\prime}$ and $\bfmu_{B} \equiv (\mu_{1},\ldots,\mu_{n_B})^{\prime}$,  we have
	\begin{equation}
	\mu(A) = \frac{1}{|A|}\int_{A}\delta(\bu)d\bu = \textbf{h}(A)^{\prime}\bfmu_{B}; \; A \subset \mathbb{R}^{2}. 
	\end{equation}

Our final process model at the aggregate level of spatial and temporal support can then be written
\begin{equation}
Y_t^{(\ell)}(A) = \bh(A)' \bfmu_B + \bfpsi_{t}^{(\ell)}(A)' \bfeta + \xi_t^{(\ell)}(A),
\label{eq:Yprocess}
\end{equation}
where $\bfpsi_{t}^{(\ell)}(A) \equiv (\psi_{1,t}^{(\ell)}(A),\ldots,\psi_{r,t}^{(\ell)}(A))'$ and $\psi_{j,t}^{(\ell)}(A) \equiv \frac{1}{\ell}\sum_{k=t-\ell+1}^t \psi_j(A;k)$ are the spatio-temporal aggregated basis functions with $\{\xi_t^{(\ell)}(\cdot)\}$ denoting mean zero Gaussian i.i.d errors that are independent in time and space. The model in (\ref{eq:Yprocess}) is nonstationary, nonseparable, asymmetric, and provides a way to easily model areal units $A \subset \mathbb{R}^{d}$ on different scales. {To our knowledge, there has been no such model proposed that allows for realistic spatio-temporal covariances (i.e., nonstationary, nonseparable, and asymmetric) that is flexible enough to model data defined on multiple spatial and temporal scales.}

\subsubsection{Random Effects Parameterization}

The most critical component of the process model in (\ref{eq:Yprocess}) is the $r$-dimensional random effects vector, $\bfeta \; \sim \; Gau({\mbf 0},{\mbf K})$, where $\bK$ is an $r \times r$ covariance matrix.  Although one could specify a general Bayesian covariance prior (e.g., through an inverse-Wishart distribution, modified Cholesky decomposition, etc.), such a prior does not reflect the underlying spatio-temporal dynamical process occurring at { fine resolutions. This motivates us to consider a parameterization for $\textbf{K}$ that incorporates fine-scale (e.g., the B-scale) dynamics within the model stated in Section 2.1. Specifically, we define a} ``target process'' {that \textit{is dynamical} on a single small-scale geography (the $B$-scale). Then, $\textbf{K}$ is chosen so that the $B$-scale covariances of $\{Y_{t}^{(\ell)}(\cdot)\}$ in (\ref{eq:Yprocess}) are close to the analogous covariances of the target process, which in the Gaussian setting, incorporates dynamics within $\{Y_{t}^{(\ell)}(\cdot)\}$ on the $B$-scale. This approach has important implications for modeling (temporal) multi-scale processes in that current parameterizations available for a single scale spatio-temporal process can be readily used to define a parameterization for multi-scale spatio-temporal processes.}

Let the ``target-process'' on the finest areal spatial scale be given by
\begin{equation}\label{pseudo}
Y_{t}^{*}(B) = \mu(B) + \nu_{t}(B), \hspace{10pt} t = T_L,\ldots,T_U, B \in D_{B},
\end{equation} 
where $\mu(B)$ is defined in (\ref{partition}), and $\nu_{t}(B)$ is a Gaussian random variable with mean zero where $\bfnu_{t}\equiv (\nu_{t}(B): B \in D_{B})^{\prime}$ is the associated Gaussian random vector. As discussed in \citet{cressie-wikle-book}, many realistic dynamical spatio-temporal processes can be represented by fairly simple first-order Markov models.  Thus,  assume that $\bfnu_{t}$ follows a first-order vector autoregressive model
\begin{equation}\label{dynamic}
\bfnu_{t} = \bM \bfnu_{t} + \bb_{t}, \;\; t = T_L,\ldots,T_U,
\end{equation}
where $\textbf{M}$ is a $r\times r$ real-valued propagator matrix. {We make use of what we have termed a ``Moran's I'' (MI) propagator \citep{bradleyMSTM}, which only requires knowledge of the adjacency structure of our areal spatial domain.  In addition, the MI propagator is helpful in that it can accommodate parsimonious time-varying behavior through the use of spatio-temporal covariates. Thus, we set $\textbf{M}$ equal to the MI propagator matrix (see the Appendix for the definition of MI propagator).}

 Given $\bM$, let $\bb_{t}$ be an $n_{B}$-dimensional Gaussian random vector with mean zero and precision matrix $\bfSigma_b^{-1} \equiv (1/\sigma_{K}^{2})(\textbf{I} - \textbf{A})$, where $\sigma_{K}^{2}>0$ is unknown, and $\textbf{A}$ is the adjacency matrix formed by the edges implied by $D_{B}$. Then, assuming stationarity, the VAR(1) structure suggests the following \citep[e.g., see][Chap. 6]{cressie-wikle-book}:
\begin{equation*}
\cov(\bY^*_{t+\tau},\bY^*_{t}) \equiv\bfSigma_{Y*}^{(\tau)} = \bM^\tau \bfSigma_{Y*}^{(0)},
\end{equation*}
where
\begin{equation*}
\mbox{vec}(\bfSigma_{Y*}^{(0)}) = [\bI - \bM \otimes \bM]^{-1} \mbox{vec}(\bfSigma_b).
\end{equation*}
Thus, given $\sigma^2_K$, $\bM$ and $\bfSigma_b$, one can specify all of the elements of
\begin{equation*}
\bfSigma_{y*} \equiv \var\{(\bY^{*\prime}_{T_L},\ldots,\bY^{*\prime}_{T_U})'\},
\end{equation*}
where $\bY_{t}^{*}\equiv (Y_{t}^{*}(A): B \in D_{B})^{\prime}$ for $t = T_L,\ldots,T_U$.  Finally, we let 
\begin{equation}\label{eq:targ}
\bK \equiv \underset{\textbf{C}}{\mathrm{arg\hspace{5pt}min}}\left\lbrace ||\bfSigma_{y*} - \bfPsi \textbf{C} \bfPsi^{\prime}||_{F}\right\rbrace,
\end{equation}
\noindent
where $\sigma_{K}^{2}>0$ is unknown, for generic square matrix $\textbf{G}$ we have $||\textbf{G}||_{F}^{2}\equiv \mathrm{trace}\left(\textbf{G}^{\prime}\textbf{G}\right)$ is the Frobenius norm, the space of $\textbf{C}$ in (\ref{eq:targ}) is restricted to be positive semi-definite matrices, and the $\sum n_{t}^{(1)}\times r$ matrix
\begin{equation}\label{basis_targ}
\bfPsi  \equiv ( \bfpsi_{T_L}^{(1)}(A)',\ldots, \bfpsi_{T_U}^{(1)}(A)')'.
\end{equation}
\noindent
Notice that the period $\ell$ is set equal to 1 in (\ref{basis_targ}), as the target process is defined on the finest spatio-temporal resolution. The solution to (\ref{eq:targ}) is known and easy to compute \citep[e.g., see][]{Higham,MSTMarxiv,bradleyMSTM,bradleyCAGE,cressieTheft}.

{The $r\times r$ covariance matrix \textbf{K} is extremely general and can easily be adapted to account for other spatio-temporal covariance structures in the multi-scale setting. That is, if one has a space-time covariance matrix (say $\bfSigma^{*}$) that is not readily interpretable on multiple spatial or temporal scales, then replace $\bfSigma_{y*}$ with $\bfSigma^{*}$ in (\ref{eq:targ}). This would produce a process $\{Y_{t}^{(\ell)}(\cdot)\}$ in (\ref{eq:Yprocess}), that has a space-time covariance function that is close (in Frobenius norm) to $\bfSigma^{*}$ on the scale in which $\bfSigma^{*}$ is defined.}

In summary, given the spatio-temporal aggregated basis functions, a MI propagator matrix, fine-scale adjacency matrix, and a single variance parameter, we are able to obtain in (\ref{eq:targ}) a positive definite marginal covariance matrix for our spatio-temporal random effects, $\bfeta$, that respect a target dynamical process at the fine (aggregated) spatial scale.  It is important to note that we do not predict $\bfnu_t$ nor $\bY^*_t$, they are just an intermediate step toward obtaining the target marginal covariance in order to apply the \citet{Higham} best positive approximate procedure. 

\subsection{Parameter and Basis Function Specification}

The model specified above is {\it extremely} parsimonious for a spatio-temporal model.  In particular, we need only specify $\sigma^2_{\xi}$, $\sigma^2_K$, and $\bfmu_B$.  In the application presented in Section 3, we assign both $\sigma^2_\xi$ and $\sigma^2_K$ a $IG(1,1)$ (inverse-gamma) prior and the elements of $\bfmu$ are assumed {\it a priori} to be i.i.d. N(0,$\sigma^2_\mu$), with $\sigma^2_\mu$ assigned an $IG(1,1)$ prior. 

The primary reason that our model can easily adapt to different levels of spatio-temporal support is because we model the spatio-temporal variability through a basis expansion. In particular, we utilize spatio-temporal basis functions and common random effects.  In this regard, it is important to specify a basis set that is flexible and can easily be integrated/aggregated in space/time.  There are many such choices in the literature \citep{wikleHandbook,bradley2014_comp}.  Here, we consider local bisquare basis functions given by
	\begin{equation}\label{cht3.bi}
	\psi_{j}(\bu;t) \equiv
	\left\{
		\begin{array}{ll}
			\{1 - (||\bu - \textbf{c}_{j}||/w_{s})^{2} - (|t-g_{t}|/w_{t})^{2}\}^{2},  & \mbox{if } ||\bu - \textbf{c}_{j}|| \le w_{s},  |t-g_{t}| \le w_{t}\\
			0,  & \mathrm{otherwise};\hspace{5pt} \bu \in D_{s},
		\end{array}
	\right.
	\end{equation}
	\noindent
	 with $j = 1,\ldots,r$ spatial knot points $\textbf{c}_{j}$, $m_t$ equally spaced temporal knot points, $g_t$, where $w_s$ and $w_t$ are maximal spatial and temporal supports. The placement of knots is chosen using a space filling design \citep{NychkaSpaceFill}. Additionally, direct Monte Carlo sampling is used to approximate ${\psi}_{j}(A; k)$. Specifically, $h$ points $\{\bs_{q}: q = 1,\ldots,h\} \subset A$ are randomly selected using a uniform distribution on $A$. Then, ${\psi}_{j}(A; k)$ is approximated with $(1/{h}) \sum_{q = 1}^{h}|A|\psi_{j}(\bs_{q}; k)$, where $\psi_{j}(\bs_{q}; k)$ is computed according to (\ref{cht3.bi}).  {In the ACS analysis considered in Section 3, we specify the components of the basis functions (i.e., $w_{s}$, $w_{t}$, and $r$ in (\ref{cht3.bi})). }

\section{ACS Median Household Income Example}
The US Census Bureau has amassed a large number of ACS period estimates, and has become an extremely rich data source. In fact, a recent Google Scholar search (on \date{7/30/2015}) of ``American Community Survey'' resulted in 3,690,000 entries. Thus, providing ACS users the flexibility to undergo space-time COS is likely to have a large impact. {To demonstrate this, in Section 3.1 we perform space-time COS to show that one can provide estimates of the 3-year period median household income using the available 1-, 3-, and 5-year period estimates. This is especially notable considering that ACS has decided to discontinue the 3-year period estimates \citep{3yearperiod}. Additionally, we provide an example of simultaneous spatial and temporal COS in Section 3.2.}

\subsection{Estimating 3-Year Period Estimates of Median Household Income}
        
We consider using ACS period estimates of median household income defined at the county level. Modeling at the county level for this illustration is partly based on the broad recognizability of this level of geography to both those within and outside of federal statistics. Also, the spatio-temporal coverage of ACS period estimates defined on counties is comparatively better than other more spatially sparse geographies; for example, census tracts have no 1-year period estimates and relatively few 3-year period estimates of median household income currently available.

        \begin{figure}[h!]
        \begin{center}
        \begin{tabular}{ll}
           \includegraphics[width=8.5cm,height=4.5cm]{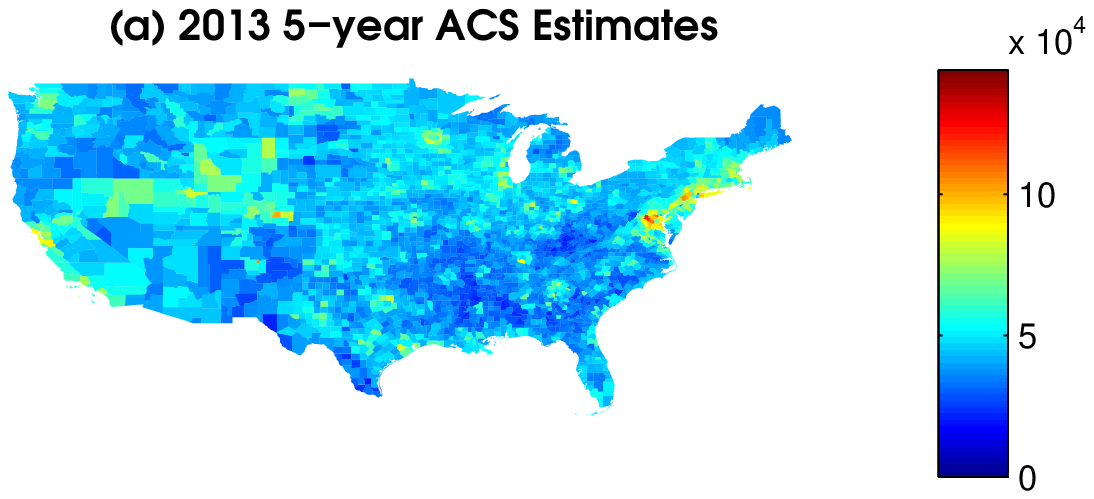}&
           \includegraphics[width=8.5cm,height=4.5cm]{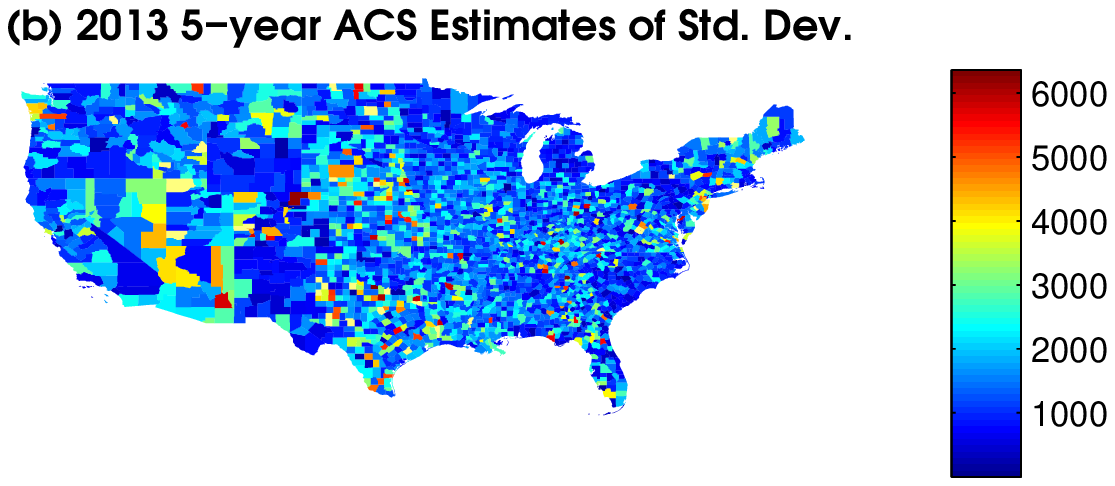}\\
                      \includegraphics[width=8.5cm,height=4.5cm]{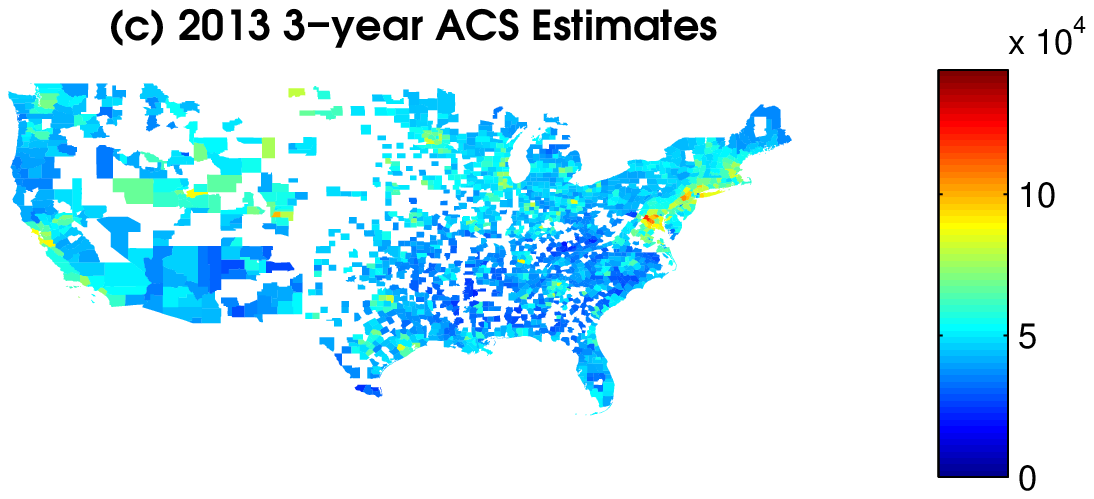}&
                      \includegraphics[width=8.5cm,height=4.5cm]{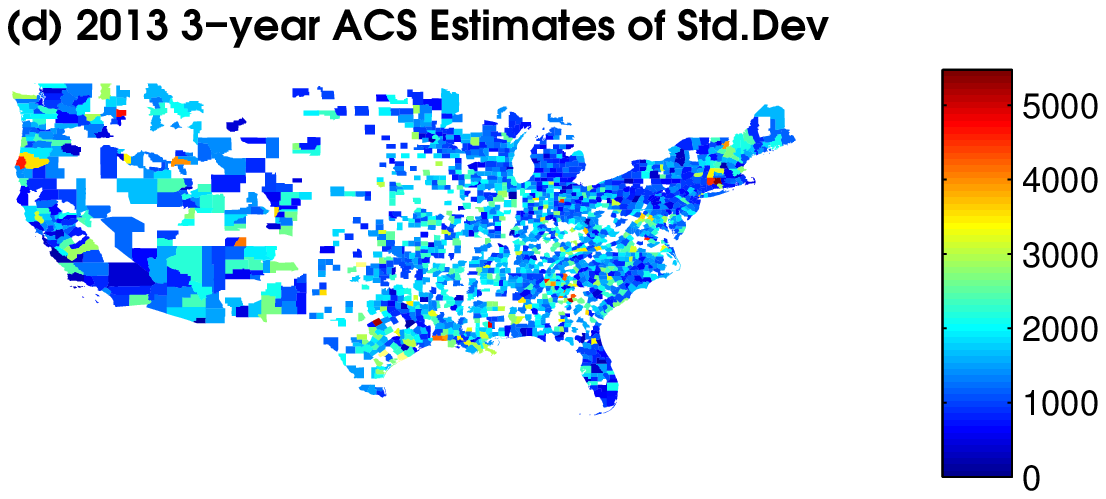}\\
                                            \includegraphics[width=8.5cm,height=4.5cm]{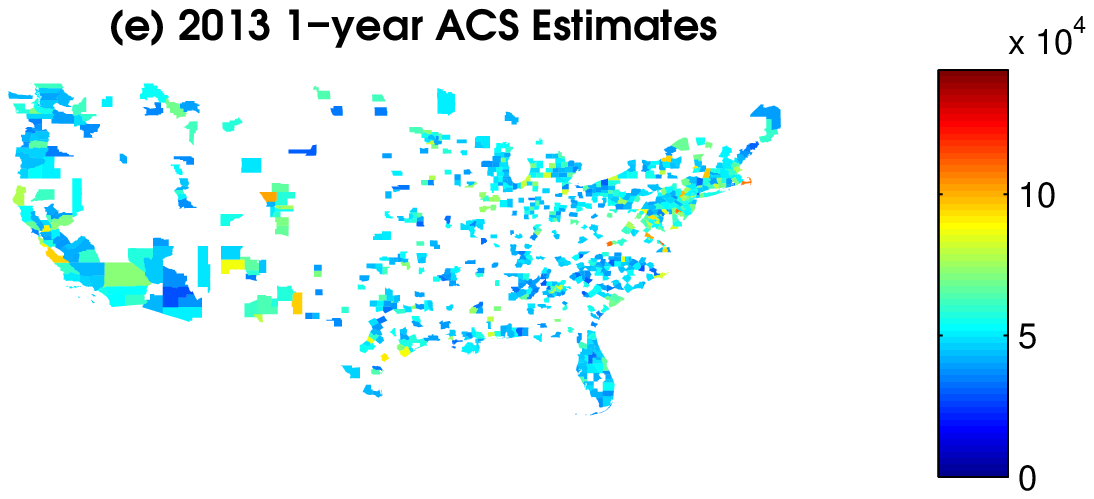}&
                                            \includegraphics[width=8.5cm,height=4.5cm]{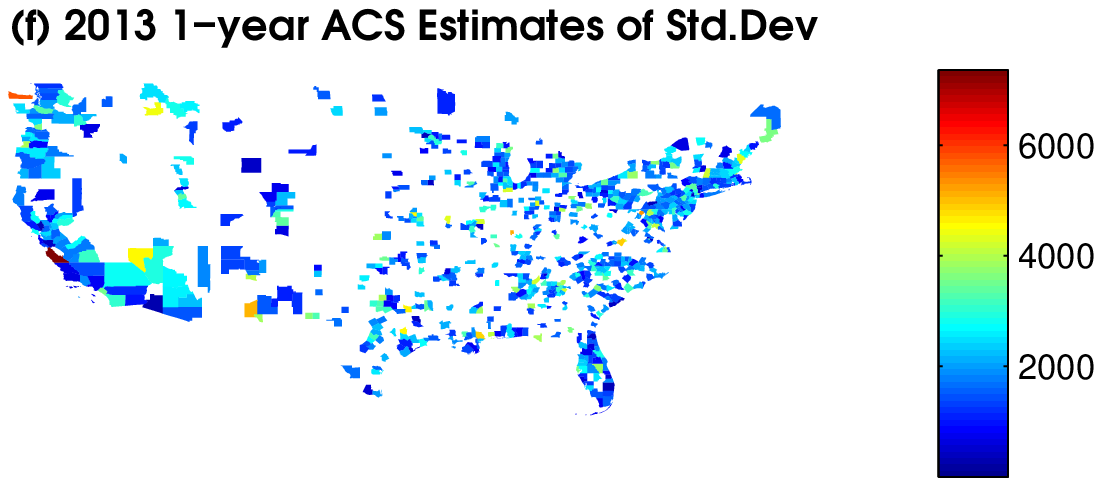}\\
           \includegraphics[width=8.5cm,height=4.5cm]{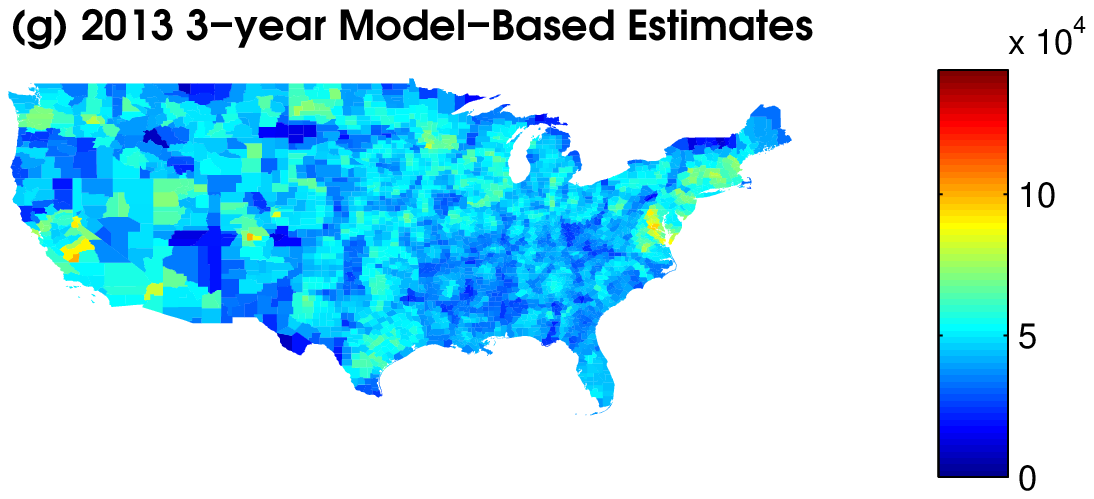}&
           \includegraphics[width=8.5cm,height=4.5cm]{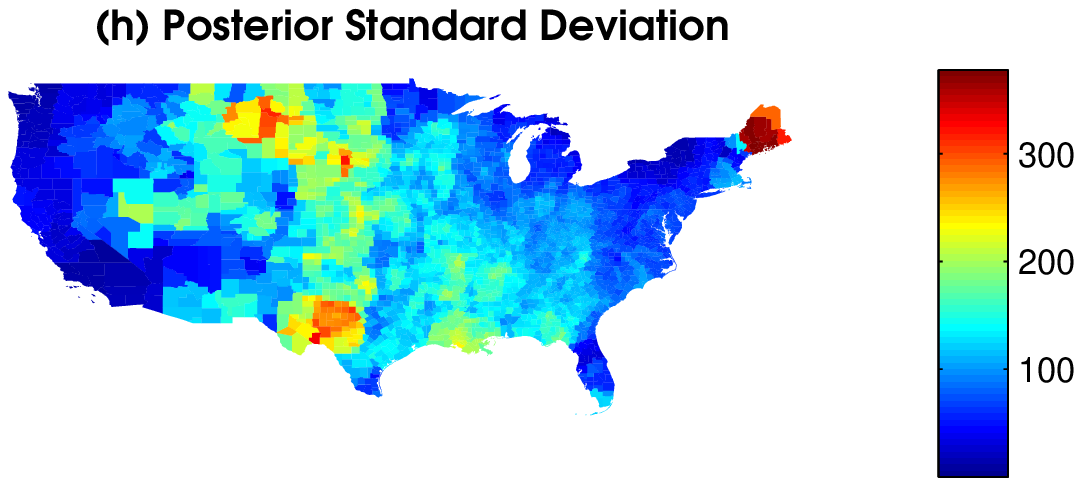}\\
        \end{tabular}
        \end{center}
        \centering
        \caption{\baselineskip=10pt {The first three rows present 2013 ACS period estimates of median household income and their respective standard deviations. The period is indicated in the title. White locations are missing. The last row presents our model based 3-year period estimate of median household income over all counties along with the associated posterior standard deviations.}}
        \label{fig:data_post}
        \end{figure}

We consider 1-year ACS estimates of median household income for 2006; 1-, 3-year period estimates for years 2007 and 2008; 1-, 3-, and 5-year estimates for 2009 through 2012; and 1- and 5-year estimates for 2013.  We exclude the 3-year period estimates for 2013 in the model fitting and estimation stage in order to keep them as a hold-out sample. To summarize, we define the data for the periods
\begin{align}
& Z_{t}^{(\ell)}(A):A \in D_{t,A}^{(\ell)},\hspace{5pt}(t,\ell) = (2006,1),\ldots,(2013,1),(2007,3),\ldots,(2012,3)(2009,5),\ldots,(2013,5).
\label{eq:Zdata}
\end{align} 
\noindent
In total, there are 19 time periods considered here, which gives a large dataset of 32,836 ACS period estimates.  As an example, Figure 1 shows the 2013 1-, 3-, and 5-year estimates of median household income along with the associated standard deviations (based on the ACS margin-of-error estimates).

  \begin{figure}[h!]
       \begin{center}
      \begin{tabular}{c}
     \includegraphics[height=12cm]{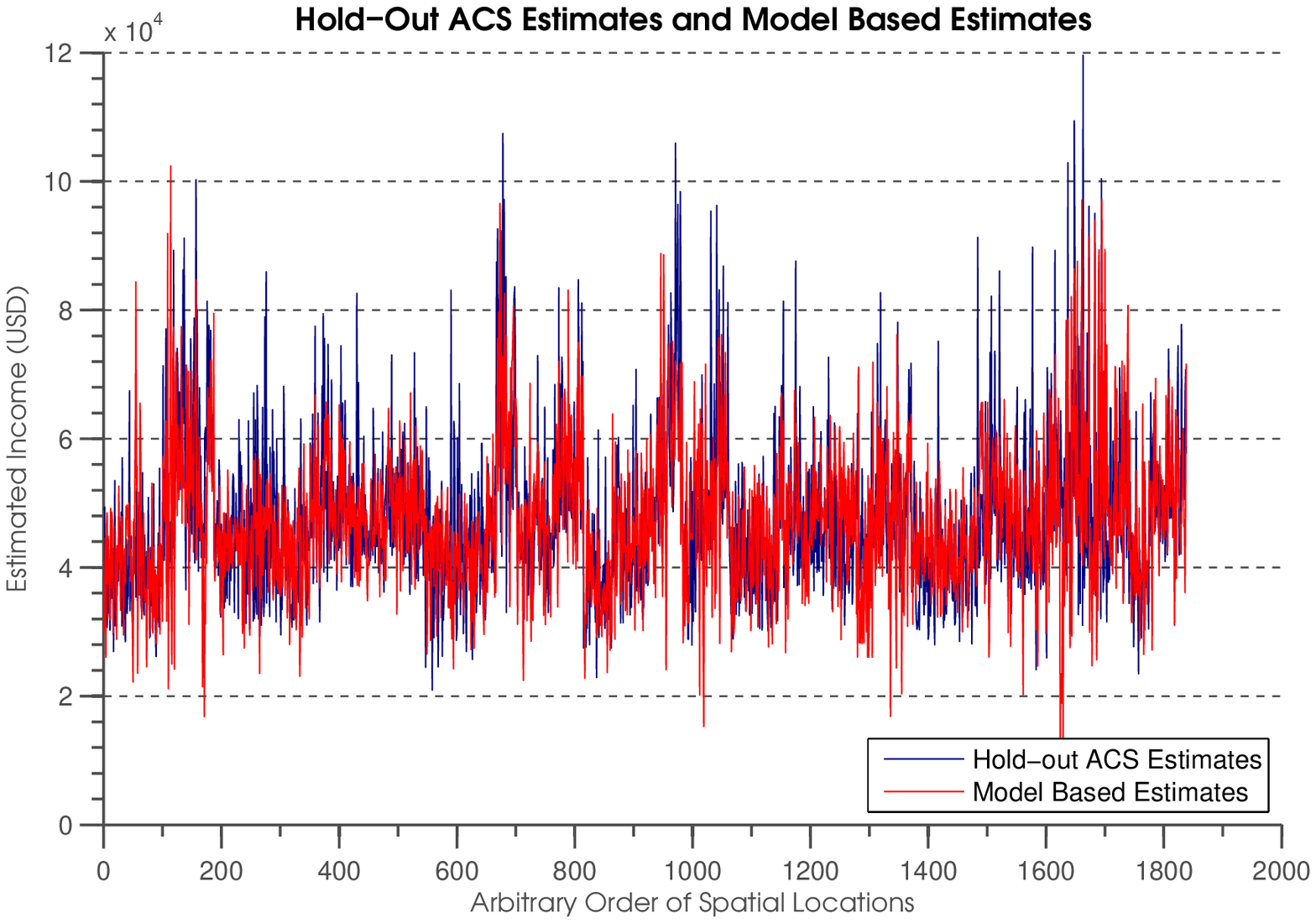}
        \end{tabular}
        \caption{\baselineskip=10pt {Predictions based on all data except $\{Z_{2013}^{(3)}(A)\}$ is given by the black lines. The estimates in the set $\{Z_{2013}^{(3)}(A)\}$ are given by the red lines.}}
        \end{center}  
        \label{fig:holdoutplot}
  \end{figure}
To compute the predictor 
\begin{equation*}
\widehat{Y}_{t}^{(\ell)}(A) \equiv E\left[Y_{t}^{(\ell)}(A)\vert \{Z_{t}^{(\ell)}(A)\}\right];\hspace{10pt} A \in D_{t,A}^{(\ell)}, \hspace{5pt}t = T_{L},\ldots,T_{U}, \hspace{5pt}\ell = 1,3,5,
\end{equation*}
where the posterior expectation is taken using the model introduced in Section 2, we need to specify the spatio-temporal basis functions. This includes defining $w_{s}$ (radius of the spatial component of the basis functions), $w_{t}$ (radius of the temporal component of the basis function), the number of knots, and the knot locations. 

	 {We set $w_{s}$ to be 1.1 times the smallest distance between two different knots, $g_{t}$ is set equal to the $m_t$ mid-points associated with the different ACS period estimates of median household income, and $w_{t}= 1.1$. The value for $w_{s}$ and $w_{t}$ were chosen to minimize hold-out error (i.e., squared distances between predictions and $\{Z_{2013}^{(3)}(\cdot)\}$). Similarly, we considered $r = 50,100,150,200,250$ and $300$ and the value of $r$ that minimizes hold-out error is $r = 250$. Thus, in the following application $r=250$ and $t=2005.5,2006,\ldots,2012.5$ (i.e., $m_t = 19$), so that $\bfpsi_{t}^{(\ell)}(A)$ is a $r \times m_t = 4750$-dimensional vector.}

 In the fourth row of Figure 1 we display the predicted median household income $Y_{2013}^{(3)}(\cdot)$. Visually, these predictions perform quite well when comparing to the hold-out ACS sample (i.e., $\{Z_{2013}^{(3)}(\cdot)\}$.  Also, notice that the posterior standard deviation is \textit{considerably} smaller than the ACS standard deviation (the model based errors are close to 20 times smaller than the ACS errors). Figure 2 provides a second look at the hold-out versus the model-based predictions, and again, our predictions appear to be performing quite well.

As a diagnostic measure, consider the ratio
 \begin{equation}\label{ratio}
 R(A)\equiv \frac{Z_{2013}^{(3)}(A)}{\widehat{Y}_{2013}^{(3)}(A)};\hspace{5pt} A \in D_{2013,A}^{(3)}.
 \end{equation}
 \noindent
Here, if $R$ is close to 1 then the model-based estimate is similar to the hold-out ACS estimate, if $R$ is less than 1 then the model-based estimate is larger than the hold-out ACS estimate, and if $R$ is greater than 1 then the model-based estimate is smaller than the hold-out ACS estimate. In Figure 3(a), we plot a histogram of the values in the set $\{R(A)\}$. The histogram is slightly skewed right, indicating that there is more of a tendency to under-predict than over-predict; however, a majority of the mass of the histogram is located at 1 {(similar behavior can be seen in Figure 2)}. Thus, we see that we are consistently reproducing similar estimates to the ACS hold-out 3-year period estimates. Finally, to further corroborate the results using the ratio, in Figures 3(b) and (c), we plot the histogram of $R(A)$ associated with the hold-out experiment that sets aside $Z_{2013}^{(3)}(\cdot)$ and $Z_{2012}^{(3)}(\cdot)$, and obtain similar results.
 
        \begin{figure}[h!]
        \begin{center}
        \begin{tabular}{cc}
           \includegraphics[width=8.5cm,height=8cm]{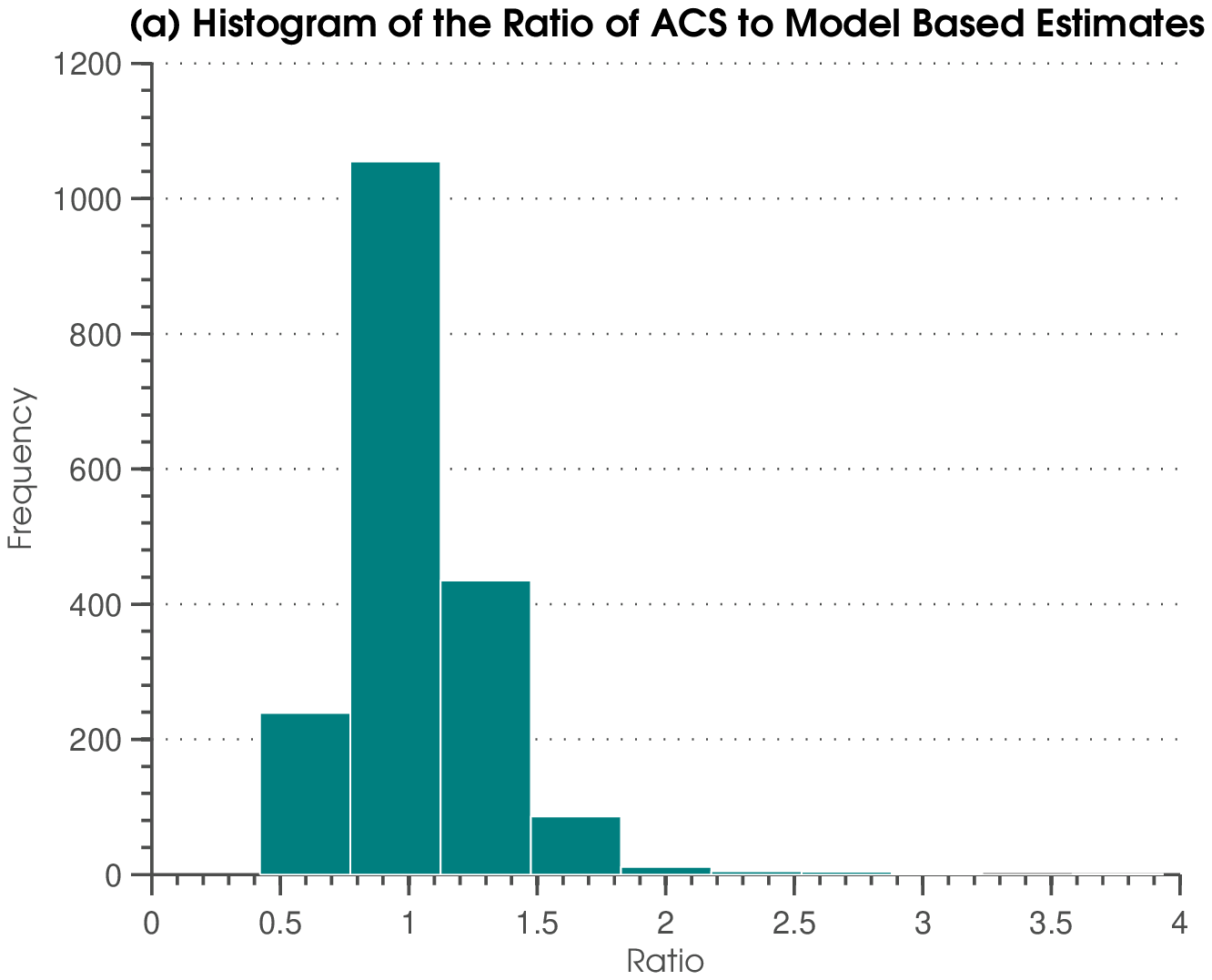}&
           \includegraphics[width=8.5cm,height=8cm]{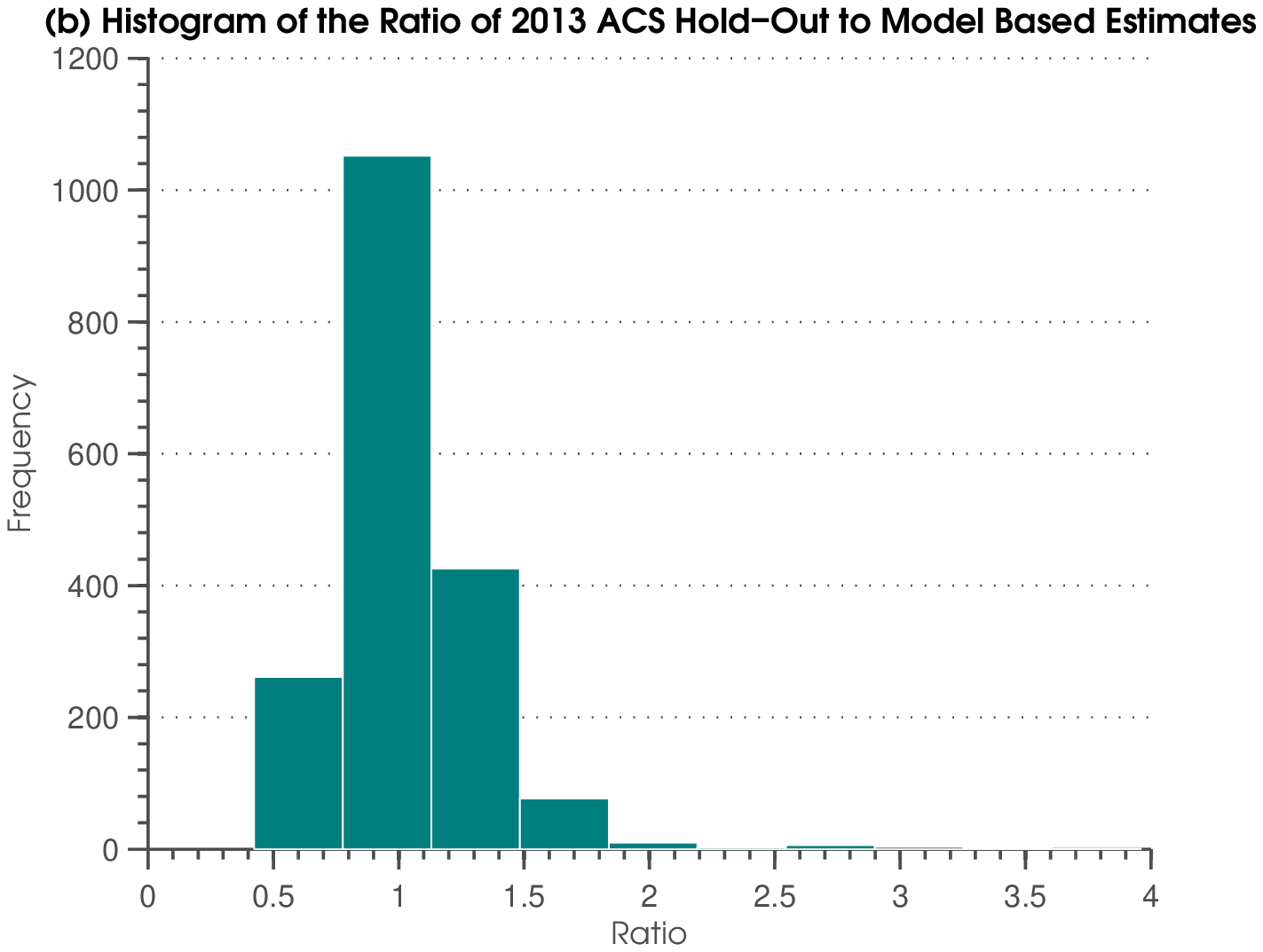}\\
     \includegraphics[width=8.5cm,height=8cm]{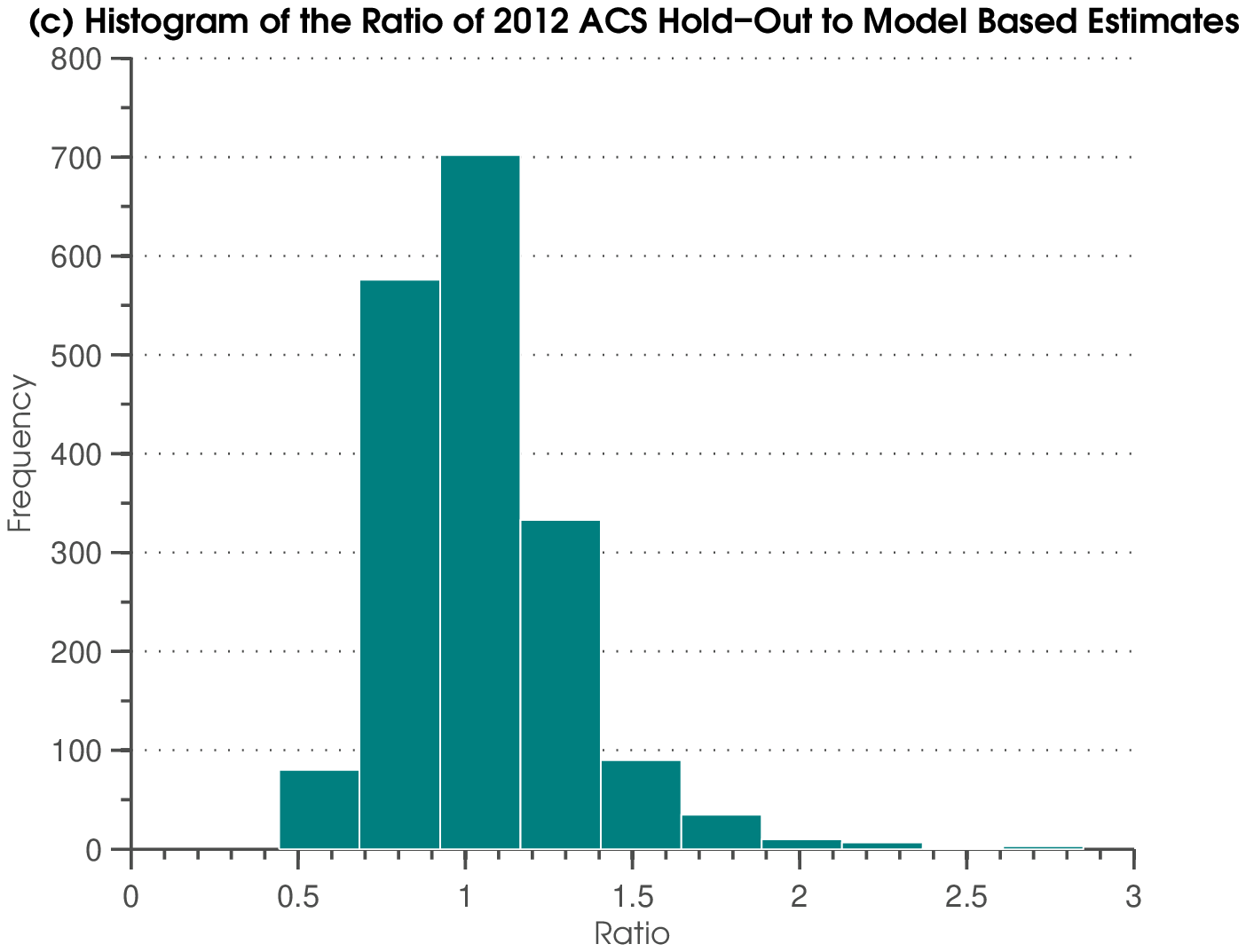}& \includegraphics[width=8.5cm,height=8cm]{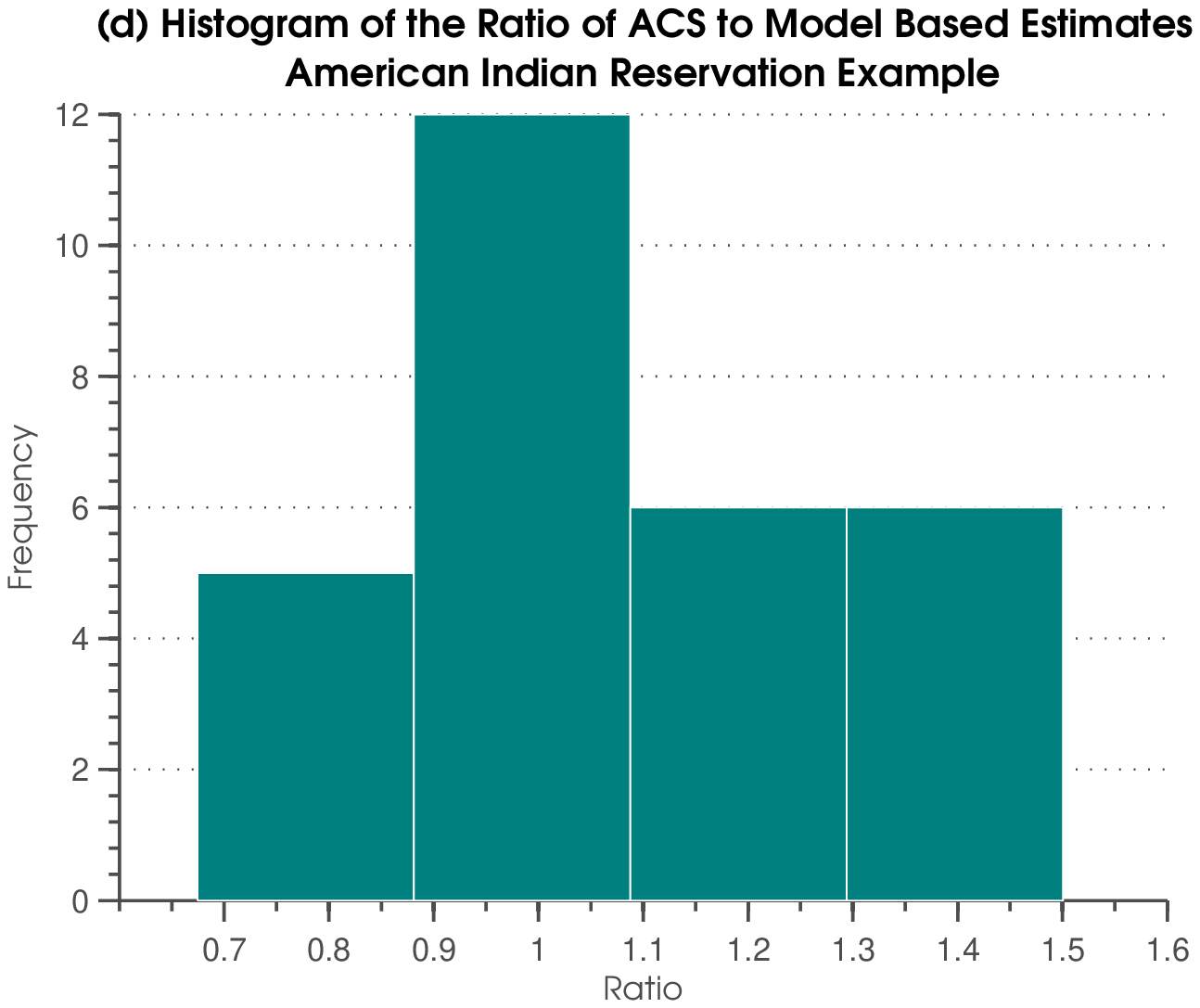}
        \end{tabular}
        \caption{\baselineskip=10pt {The ratio in (5). Panel (a) gives a histogram of the ratio between $\{Z_{2013}^{(3)}(A)\}$ and the predictions that hold out $\{Z_{2013}^{(3)}(A)\}$. Panel (b) gives a histogram of the ratio between $\{Z_{2013}^{(3)}(A)\}$ and the predictions that hold out both $\{Z_{2013}^{(3)}(A)\}$ and $\{Z_{2012}^{(3)}(A)\}$. Panel (c) gives a histogram of the ratio between $\{Z_{2012}^{(3)}(A)\}$ and the predictions that hold out both $\{Z_{2013}^{(3)}(A)\}$ and $\{Z_{2012}^{(3)}(A)\}$. {Panel (d) gives a histogram of the ratio between the 2013 3-year period ACS estimates of median income over American Indian area/Alaska native area/Hawaiian home lands, and the corresponding model-based estimates (see Section 3.2)}.}}
        \end{center}
        \label{fig:hist}
        \end{figure}

\subsection{Example of Simultaneous Spatial and Temporal COS} {In 1975 Congress appended Section 203 to the Voting Rights Act, which provides voting resources for US citizens that are not proficient in English. Recently, \citet{joyce} described a precise approach to classify regions that satisfy the jurisdiction rule laid out by Section 203. Their results indicated that many American Indian areas/Alaska native areas/Hawaiian home lands, met the jurisdiction rule of Section 203. Thus, considering the need for language assistance among US citizens that reside in these areas, it would be worthwhile to determine whether additional assistance is required based on (low) income status. Note that it has been shown that those at lower income levels tend to face certain obstacles (e.g., transportation needs), making it difficult to vote \citep[see][for more discussion]{gelmanvote}. We use this important example to demonstrate simultaneous spatial and temporal COS.

Using replicates from the posterior distribution in Section 3.1 we produce model-based 3-year period estimates of the median income of individuals that reside in American Indian areas/Alaska native areas/Hawaiian home lands. In Figure 3(d), we plot a histogram of the ratio in (\ref{ratio}) using ACS 3-year period estimates of median income in American Indian areas/Alaska native areas/Hawaiian home lands. As in Figure 3(a,b,c), the ratios are consistently close 1 indicating strong out-of-sample performance of our proposed method. 

In Figure 4, we provide time series plots for 3 regions in our target support: the Navajo Nation reservation and off-reservation trust land, the Uintah and Ouray reservation and off-reservation trust land, and the Wind River reservation and off-reservation trust land. This is done to highlight the ability of our method to compare across geographic regions over time (that are incompatible when using ACS data alone), which is a problem of interest among the federal statistics community \citep[e.g., see][]{tucker}. We choose these specific regions because they display the most notable patterns in median income, and in general, the time series plots associated with each region display different patterns. In Figure 4 (a,c,e,g), we see that the {Navajo Nation reservation} has low median income, and that both the {Uintah and Ouray reservation} and the {Wind River reservation} have median incomes that increase and then decrease over time. Then in Figure 4(b,d,f,h), we see that we obtain precise estimates using our approach (i.e., the posterior standard deviations are small relative to the scale of the data). Furthermore, the standard deviations appear to be larger for earlier years than for later years, which conforms to intuition since more ACS estimates are available as time goes on. Additionally, as we increase the period we see that the posterior standard deviations are smaller.}

\section{Discussion}

In its relatively short period of existence, the ACS has proven to be a valuable resource for data users across academia, government, and industry.  In many cases, data users prefer to have estimates at different geographies and/or time periods than are provided by the U.S. Census Bureau.  We present a novel approach to spatio-temporal COS that allows data users to take published ACS MYEs and provide predictions at arbitrary spatio-temporal levels of support.  Our approach is novel in that it is based on a spatio-temporal mixed-effects model that relies on a spatio-temporal basis expansion, wherein the basis functions are easily aggregated and the associated random effects are independent of scale.  In addition, we provide a novel and extremely parsimonious dynamic model motivation for the marginal covariance matrix of these random effects.  This representation allows implementation on extremely large datasets.  We illustrate the effectiveness of our methodology on a  holdout sample of 3-year ACS MYEs for 2013, analogous to a potential real-world implementation that will arise when 3-year MYEs are discontinued in 2016. {Additionally, we demonstrate simultaneous spatial and temporal COS from ACS county-level period estimates to 1-, 2-, 3-, and 4-year model based estimates of median income defined on American Indian areas/Alaska native areas/Hawaiian home lands.}

The approach proposed here has clear utility beyond the ACS application presented in Section 3, and even outside of the small area estimation context of federal statistics. Specifically, with a different choice of basis functions and possibly a different dynamic structure, this methodology can be applied in areas as diverse as environmental science, and ecological modeling among others. Nevertheless, the potential impact to data-users and policy makers interested in ACS custom-designed tabulations is unparalleled.

The approach we propose provides several opportunities for methodological extension. In particular, count data is prevalent within the ACS and other federal surveys; thus, there is scope for extending spatio-temporal COS for non-Gaussian data, as was done in the spatial only case in \citet{bradleyCOS}. Additionally, to date there is relatively little (by comparison) literature in the context of multivariate spatial and spatio-temporal COS. The methods proposed here and in \citet{bradleyMSTM} provide the initial building blocks for developing a comprehensive framework and is a subject of future research.

\section*{Appendix: Review of Moran's I (MI) Propagator}
\renewcommand{\theequation}{A.\arabic{equation}}
\setcounter{equation}{0}
We provide a review of the MI propagator matrix, which was recently introduced in \citet{bradleyMSTM}. Confounding in mixed effects models is the core motivator for the MI propagator matrix, but it has additional benefits related to model parsimony and the ability to accommodate covariate-based non-autonomous propagators. By confounding, we mean that the columns of the design matrix are linearly dependent with the columns of the coefficients of random effects. 

To see this, rewrite (\ref{pseudo}) in vector form as,
\begin{equation}\label{vectorizedMoFo}
\textbf{y}_{t}^{*} = \bH \bfmu + \bfnu_{t},
\end{equation}
where $\bH \equiv (\bh(A)^{\prime}: B \in D_{B})^{\prime}$ and $t = T_L,\ldots,T_U$. Then, substitute (\ref{dynamic}) into (\ref{vectorizedMoFo}) to obtain
\begin{equation}\label{matrix:process}
\bY_{t}^{*} = \bH \bfmu + \bM_{t}\bfmu_{t-1} + \bb_{t} ;\hspace{5pt} t = T_L,\ldots,T_U.
\end{equation}
Depending on our choice for $\{\bM_{t}\}$ there might be issues with confounding between $\bfnu_{t-1}$ and the $2n$-dimensional random vector $\bfzeta_{t} \equiv (\bfmu^{\prime}, \bb_{t}^{\prime})^{\prime}$; $t = T_L,\ldots,T_U$. Upon rewriting (\ref{matrix:process}), we get
\begin{equation}\label{matrix:process2}
\bY_{t}^{*} = \bB \bfzeta_{t} + \bM_{t} \bfnu_{t-1};\hspace{5pt} 
\end{equation}
where the $n\times 2n$ matrix $\bB \equiv (\bH, \bI)$. The strategy, is to set the columns of the propagator matrix $\bM_{t}$ equal to columns in the orthogonal complement of the column space of $\bB$. This ensures that the columns of $\bB$ are linearly independent of the columns of $\bM_{t}$. Now, using the spectral representation  of $(\bI - \bB(\bB^{\prime} \bB)^{-1}\bB^{\prime}) \textbf{W} (\bI - \bB(\bB^{\prime} \bB)^{-1}\bB^{\prime})= \bfPhi_{B}\bfLambda_{B}\bfPhi_{B}^{\prime}$ we set the $r\times r$ real matrix $\bM_{t}$ equal to the first $r$ columns of $\bfPhi_{B}$ for each $t$, which is denoted with $\bM_{B}$. Here, any $r\times r$ real-valued matrix $\textbf{W}$ can be used, and in Section 3 we set $\textbf{W} = \bI$ as is done in \citet{bradleyMSTM}.

\section*{Acknowledgments} This research was partially supported by the U.S. National Science Foundation (NSF) and the U.S. Census Bureau under NSF grant SES-1132031, funded through the NSF-Census Research Network (NCRN) program.

\bibliographystyle{jasa}  
\bibliography{myref}

        \begin{figure}[h!]
        \begin{center}
        \begin{tabular}{ll}
           \includegraphics[width=8.5cm,height=4.5cm]{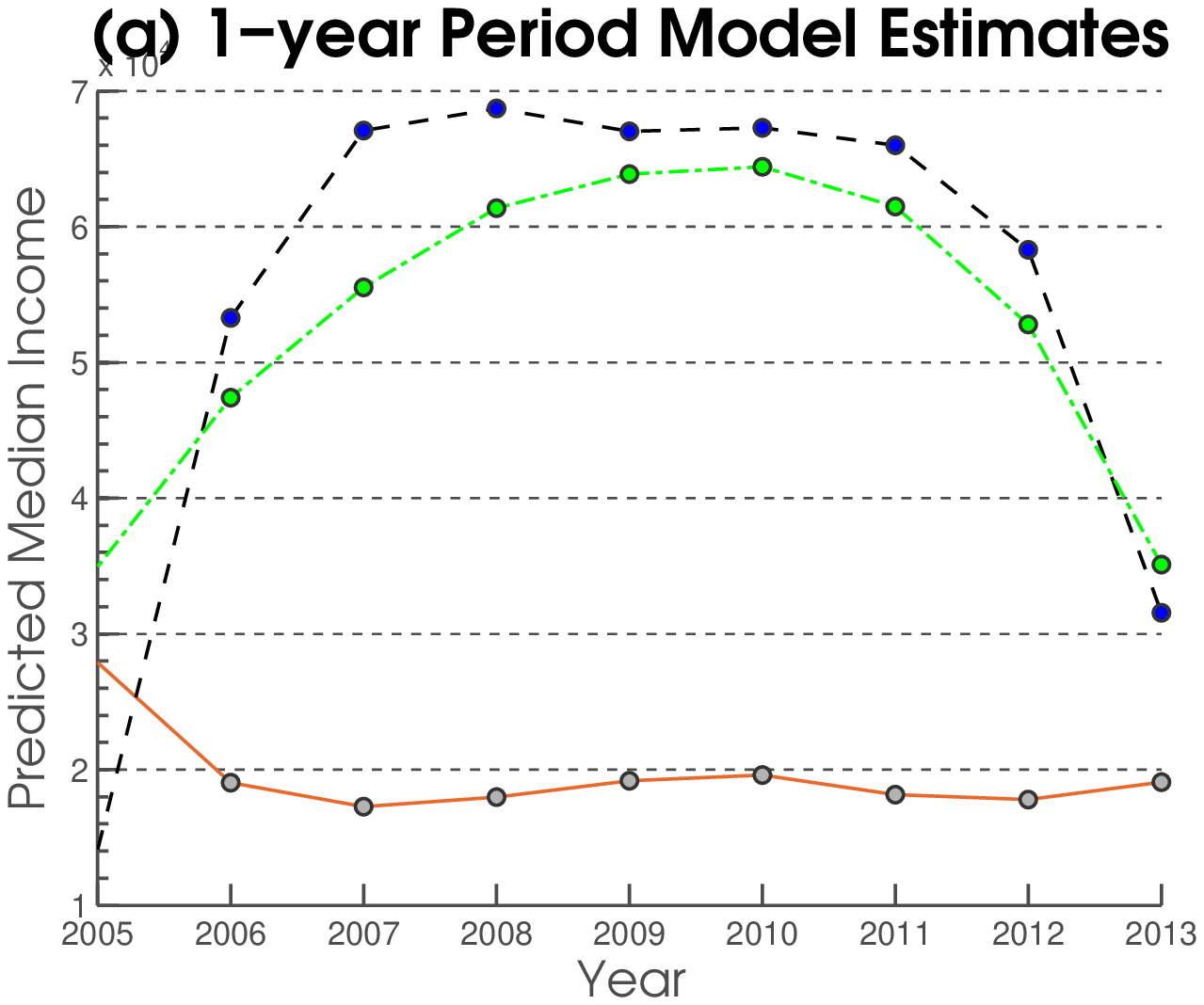}&
           \includegraphics[width=8.5cm,height=4.5cm]{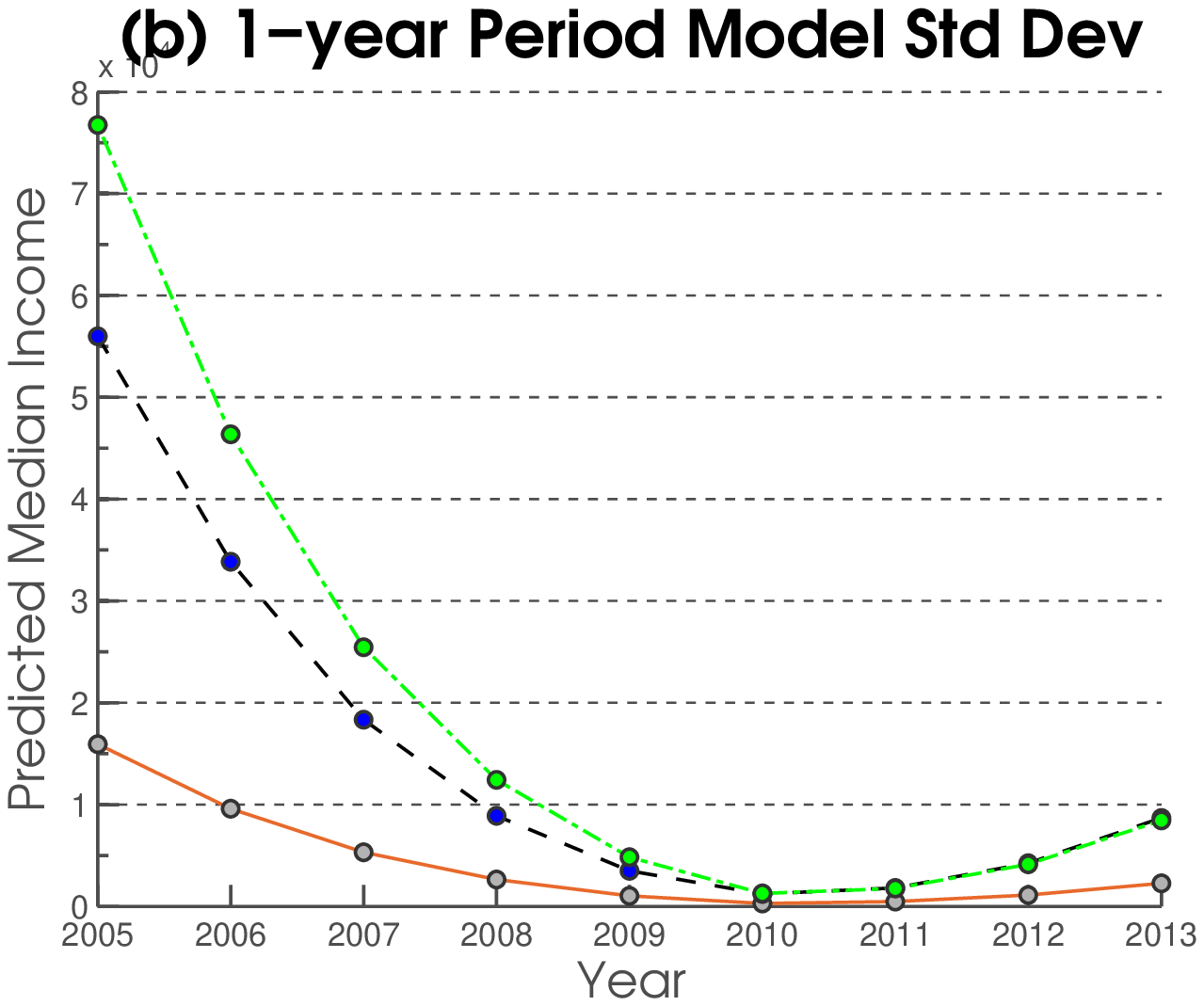}\\
                      \includegraphics[width=8.5cm,height=4.5cm]{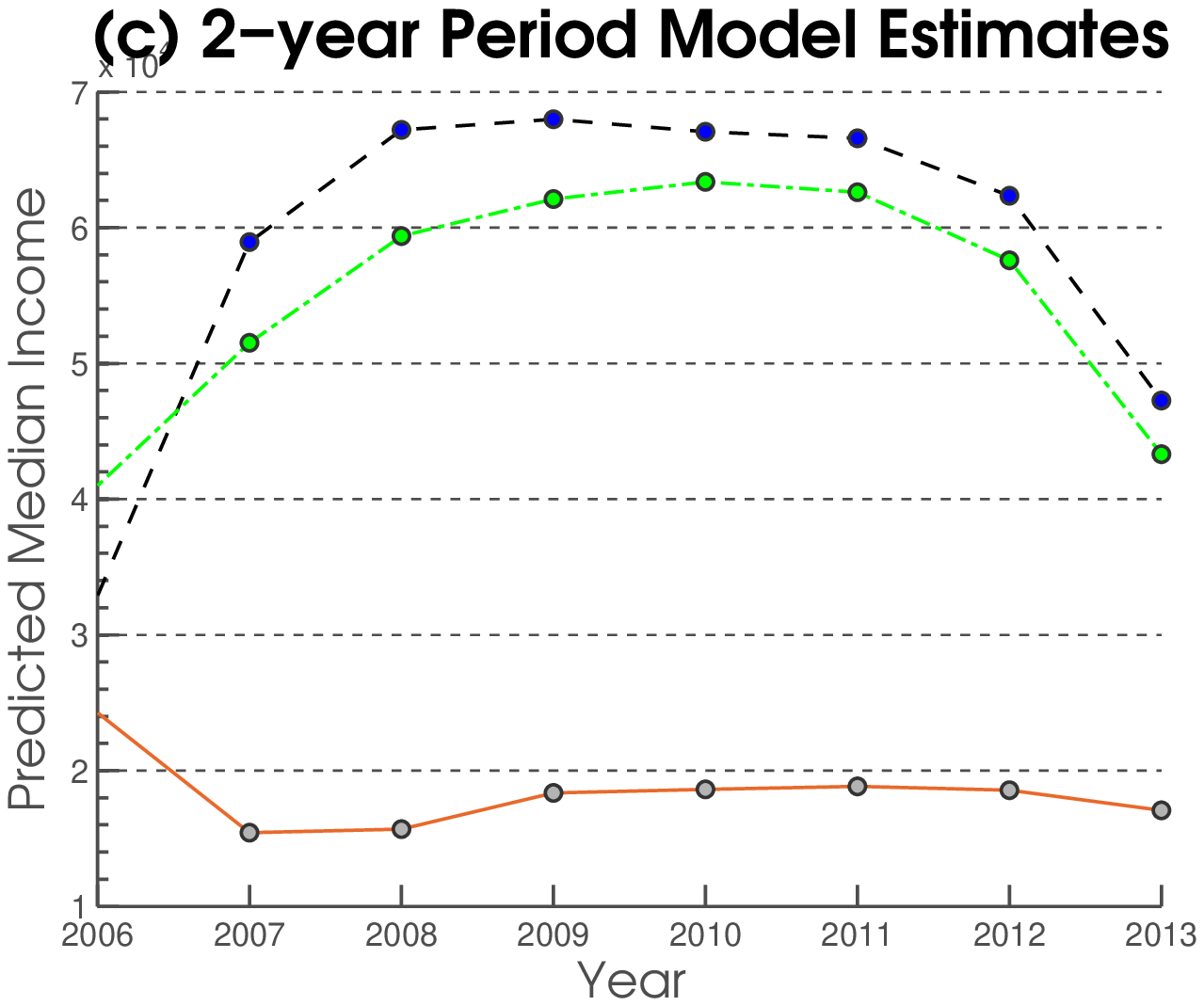}&
                      \includegraphics[width=8.5cm,height=4.5cm]{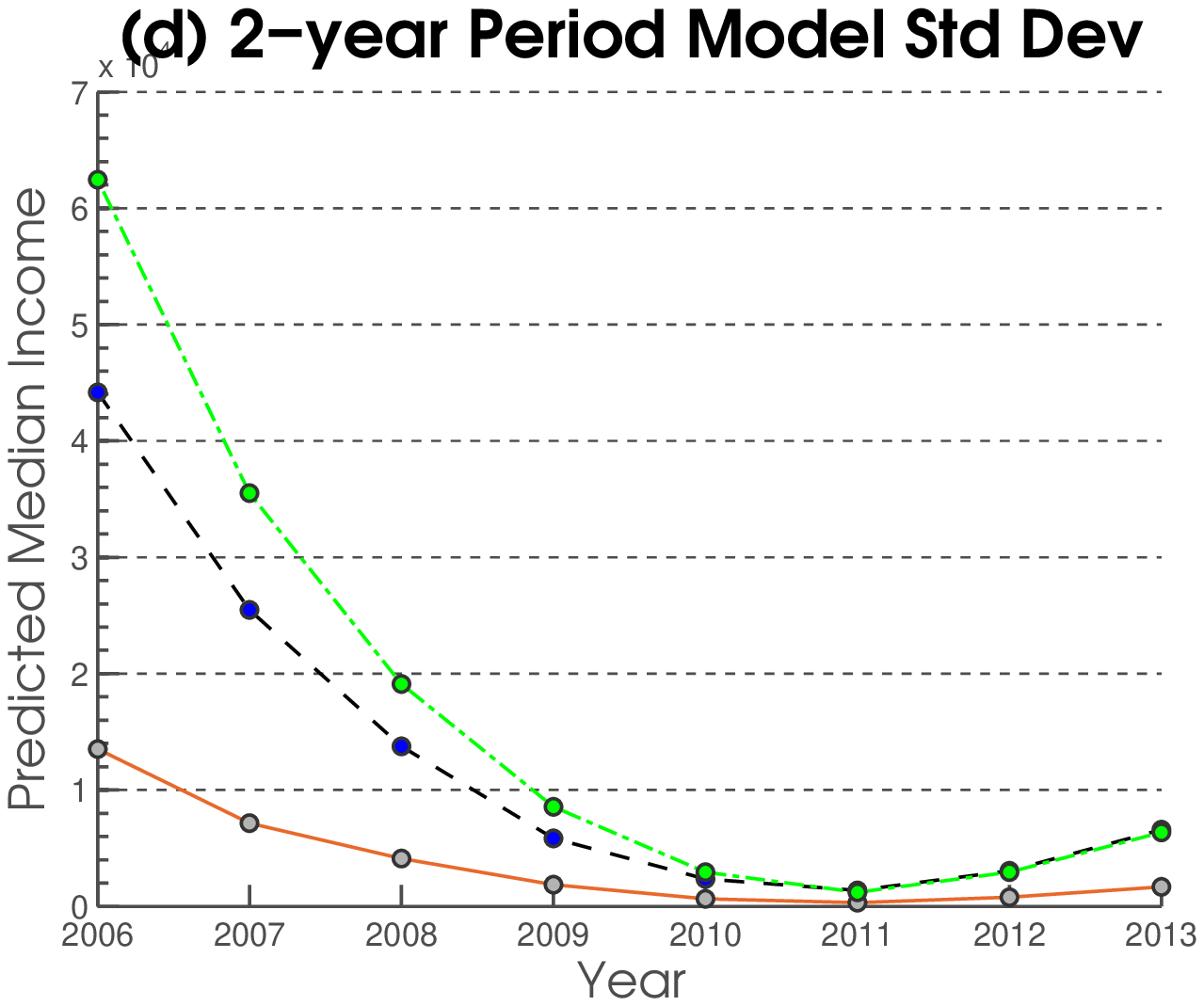}\\
                                            \includegraphics[width=8.5cm,height=4.5cm]{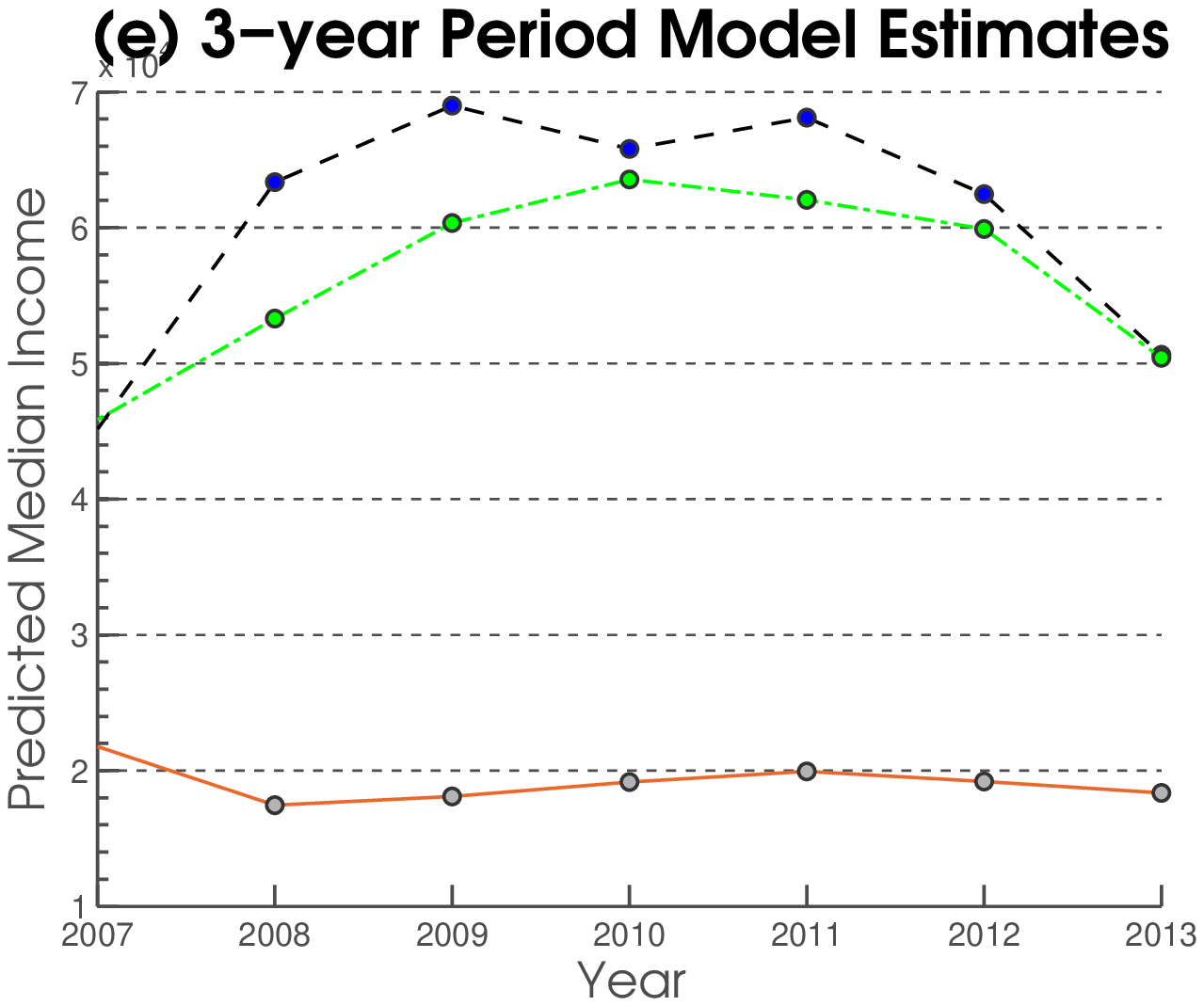}&
                                            \includegraphics[width=8.5cm,height=4.5cm]{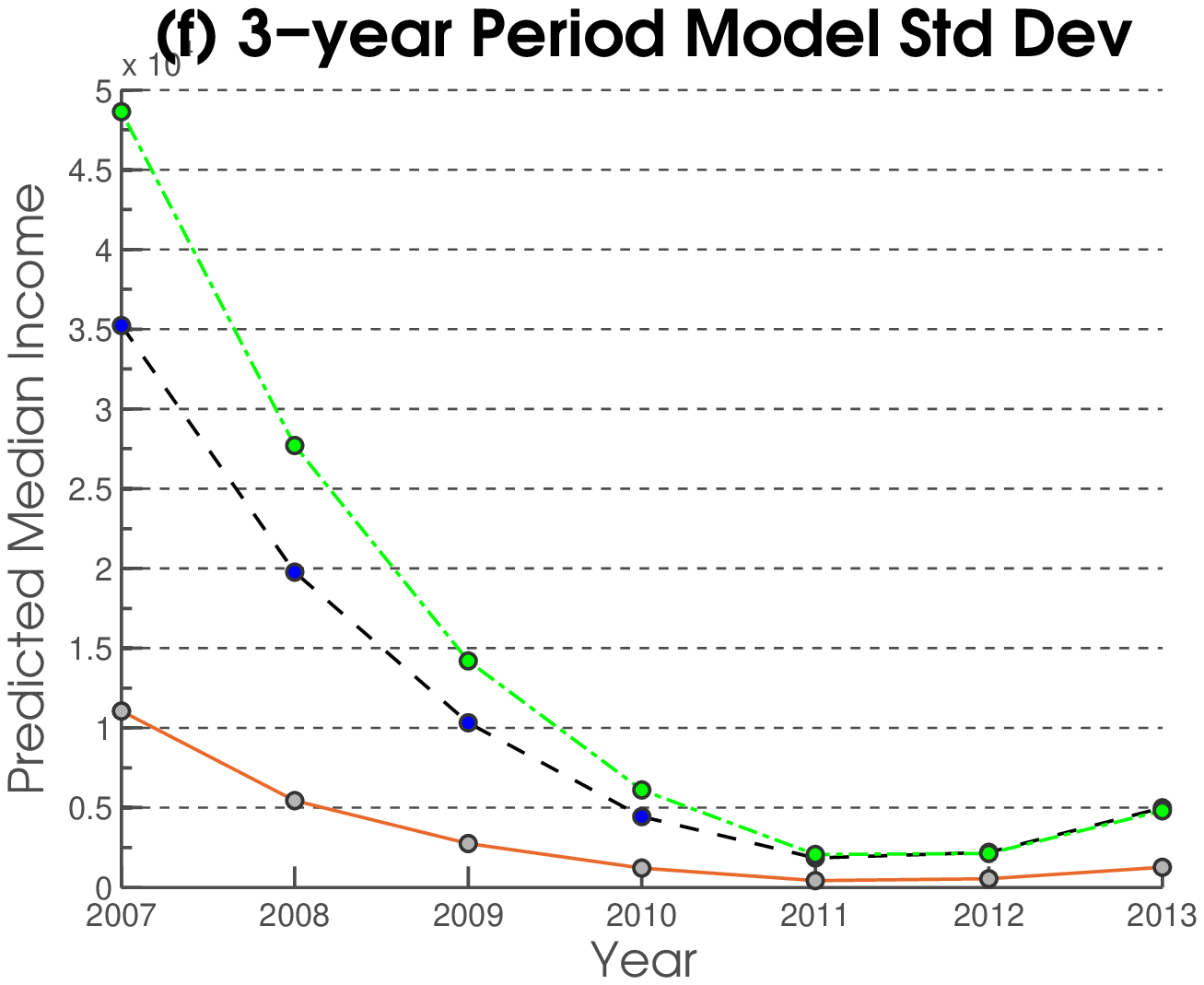}\\
           \includegraphics[width=8.5cm,height=4.5cm]{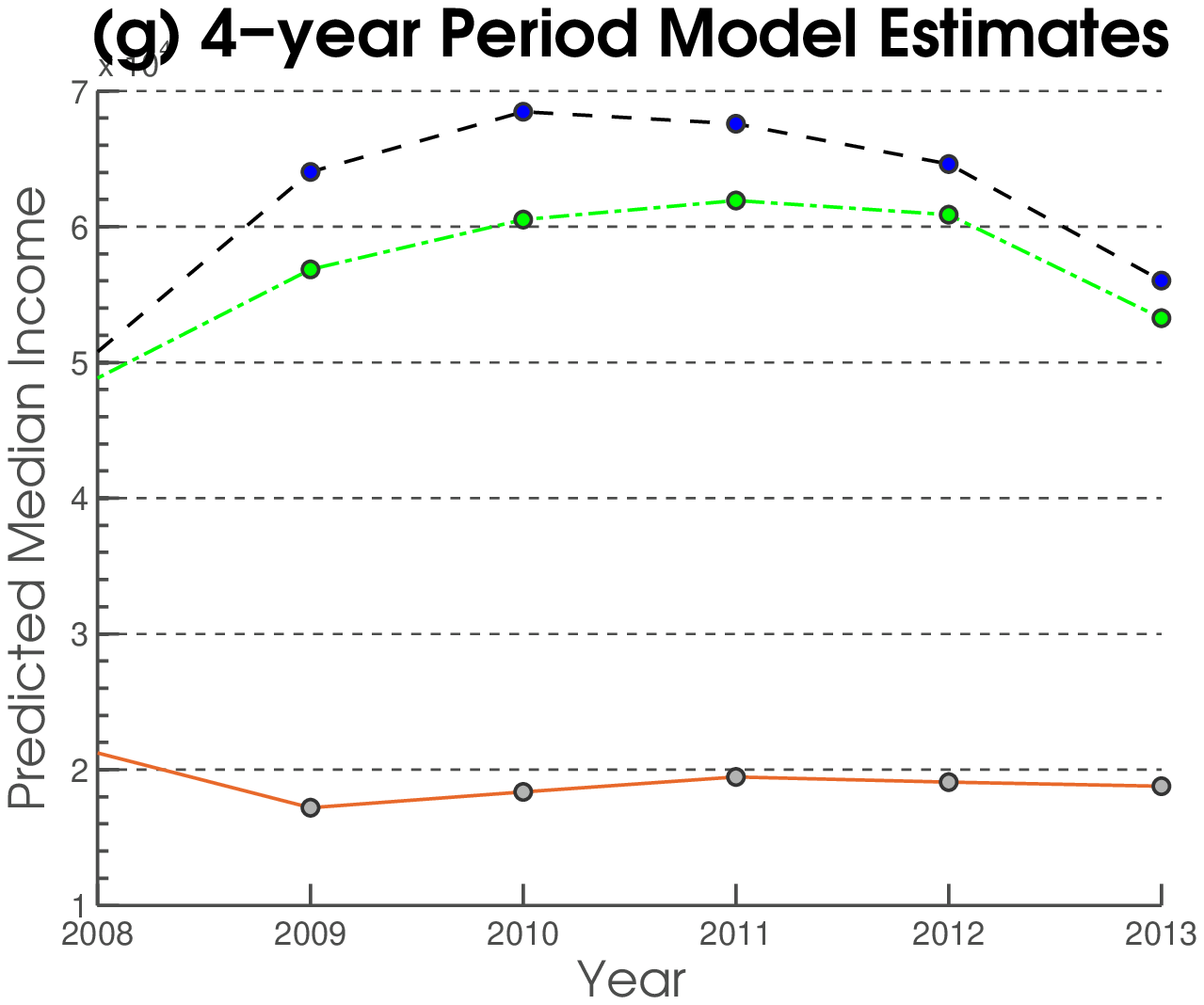}&
           \includegraphics[width=8.5cm,height=4.5cm]{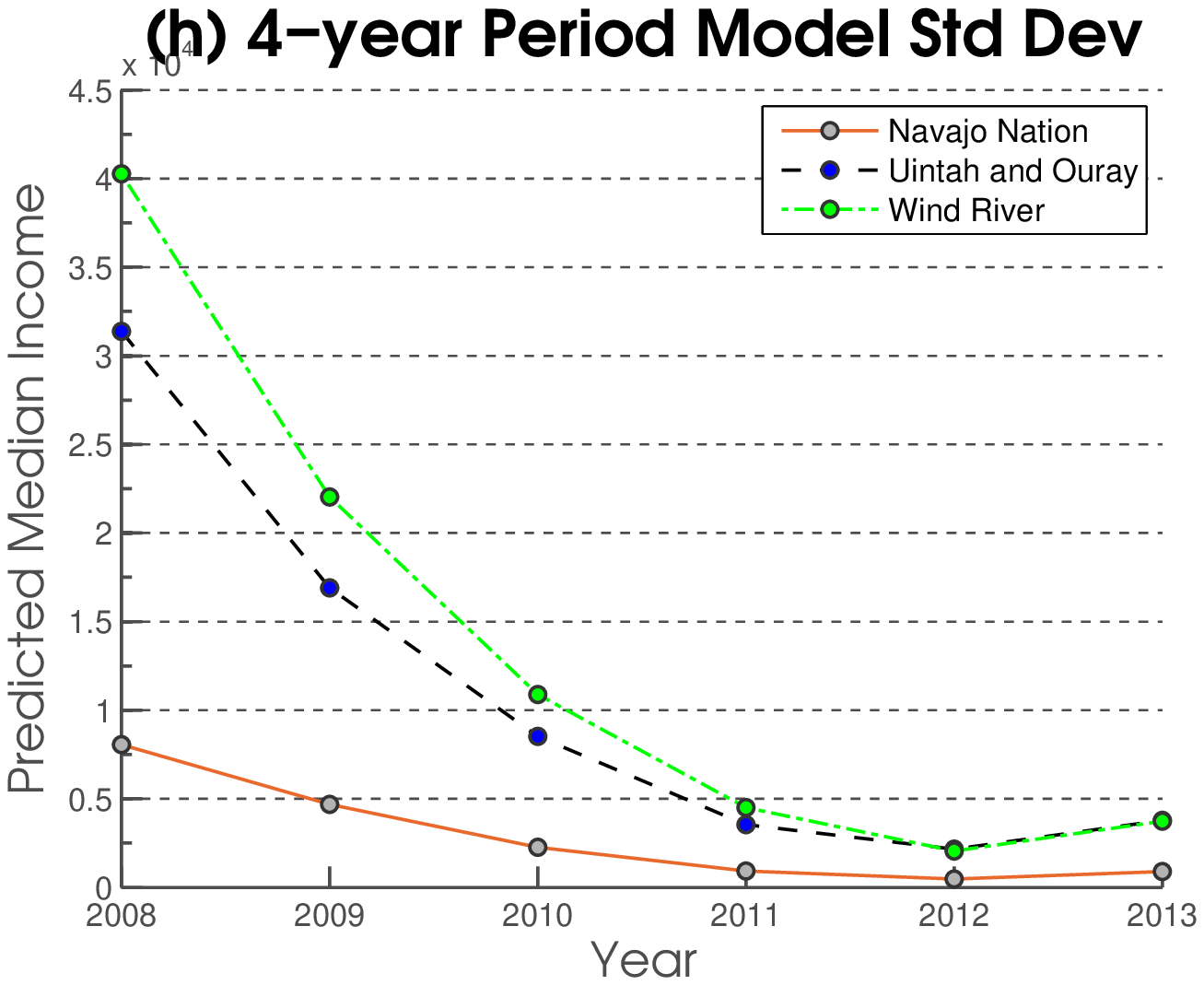}\\
        \end{tabular}
        \end{center}
        \centering
        \caption{\baselineskip=10pt {{The first column gives time series plots for 1-, 2-, 3-, and 4-year model based period estimates of median income for the Navajo Nation reservation, the {Uintah and Ouray reservation}, and the Wind River reservation, respectively. The right column displays the corresponding posterior standard deviation associated with the estimates given in the first column. The legend indicating the American Indian reservation is given in Panel (h). Notice that the each row has a different range of years indicated on the $x$-axis, and the $y$-axis differs in Panels (b), (d), (f), and (h).}}}
        \label{fig:data_post}
        \end{figure}

\end{document}